# Electrostatic co-assembly of iron oxide nanoparticles andpolymers : towards the generation of highly persistent superparamagnetic nanorods

J. Fresnais, J.-F. Berret@, B. Frka-Petesic, O. Sandre and R. Perzynski

*Matière et Systèmes Complexes, UMR 7057 CNRS Université Denis Diderot Paris-VII, Bâtiment Condorcet, 10 rue Alice Domon et Léonie Duquet, 75205 Paris, France - UPMC Univ Paris 06 - Laboratoire Liquides Ioniques et Interfaces Chargées, UMR 7612 CNRS, 4 place Jussieu-case 63, F-75252 Paris Cedex 05 France*

A paradigm proposed recently by Boal et al. (A.K. Boal *et al.*, Nature **404**, 746-748, 2000) deals with the possibility to use inorganic nanoparticles as building blocks for the design and fabrication of colloidal and supracolloidal assemblies. It is anticipated that these constructs could be made of different shapes, patterns and functionalities and could constitute the components of future nanodevices including sensors, actuators or nanocircuits. Here we report a protocol that allowed us to fabricate such nanoparticle aggregates. The building blocks of the constructs were anionically coated iron oxide nanocrytals (superparamagnetic, size 7 nm) and cationic-neutral block copolymers. We have shown that the electrostatic interactions between charged species can be controlled by tuning the ionic strength of the dispersion. Under appropriate conditions, the control of electrostatics resulted in the elaboration of spherical or elongated aggregates at the micrometer length scale. The elongated aggregates were found to be rod-like, with diameters of a few hundred nanometers and lengths between 1 and 50 µm. In addition to their remarkable stiffness, the nanostructured rods were found to reorient along with an externally applied magnetic field, in agreement with the laws of superparamagnetism.

Inorganic nanoparticles made from gold, metal oxides or semiconductors are emerging as the central constituents of future nanotechnological developments. Interest stems from the combination of complementary attributes, such as a size in the nanometer range and unique physical features including magnetic and optical properties. Among the wide variety of inorganic particles available, magnetic nanoparticles (MNP) have attracted much attention for their potential applications in materials science and biomedecine [1,2]. Dispersed in biological environments, magnetic nanoparticles exhibit the capability to "get close" to a biological entity of interest, and at the same time to be visualized in Magnetic Resonance Imaging through the modification of the water proton relaxation time. Their specific accumulation in cells or tissues permits, in a non invasive way a visualization of suspected locations and organs. Other applications of MNPs involve cell separation, diagnosis and therapy [1]. In response to these prospects, the functionalization of MNPs as well as the study of their behavior in complex environments have become key issues during the last years.

More recently, a new paradigm has gained momentum in the interdisciplinary fields of chemistry, physics and biology. This paradigm deals with the possibility to use inorganic nanoparticles as building blocks or "atoms" for the design and fabrication of colloidal and supracolloidal assemblies, or "molecules". It was suggested that these constructs could be made with different shapes, patterns and functionalities, and as such that they would cover the range that spans from the nanometer to the micrometer at once. Almost ten years ago, Boal and coworkers have shown in a pioneering work [3] the feasibility of the "brick and mortar" approach using gold nanoparticles [4]. Functionalized by hydrogen bonding self-assembled monolayers, the size and morphology of submicrometric clusters were controlled thermally, yielding spherical structures with aggregation numbers in the million range. Since then, other scenarios of controlled clustering of particles have been elaborated [5-11]. Concerning 3D-assemblies, the realization of "colloidal molecules" reminding the spatial regularity of chemical bonds between atoms was made possible using emulsions droplets filled up with colloidal particles [12] or by the control of the density of polymerization nucleation sites at the surface of inorganic colloids [13]. Glotzer and Solomon [14] have reviewed the engineering of anisotropic constructs built on the paradigm of "bricks and mortar" suggested by Boal and coworkers [3,4]. Design rules for assembly through anisotropy were posed, including patchiness, faceting, branching and chemical ordering of the building blocks. Anisotropic nanostructures, organized either in one or two dimensions have been identified as the elementary components of future nanodevices, including sensors, actuators and nanocircuits [11,15,16].

Figure 1 describes the protocol that controlled the nanoparticle co-assembly, with and without magnetic field. Solutions of poly(trimethylammonium ethylacrylate methylsulfate-b-poly(acrylamide) (VIAL **a**) and of $PAA_{2K}$-coated iron oxide nanoparticles (VIAL **b**) were prepared in 1 M ammonium chloride ($NH_4Cl$) at weight concentration c = 0.2 wt. %. At this salt content, the Debye length is of the order of 0.3 nm and electrostatic interactions are screened. It was verified by dynamic light scattering that the colloidal stability of the initial dispersions was not disrupted by the presence of a high ammonium chloride content ($D_H$ = 11 ± 1 nm for the polymer and 19 ± 1 nm for the coated MNPs). The two solutions were then mixed, yielding a disperse solution where polymers and particles were not yet associated ($D_H$ = 21 ± 1 nm, VIAL **c**). The relative amount of each component was monitored by the ratio of the electrostatic charges borne by the particles and by the polymers. This ratio was fixed at the charge stoichiometry [17]. The electrostatic interaction between the oppositely charged species was then monitored by a slow removal of the salt, either by dialysis or by dilution. Here, we discuss the last step of the protocol in Figure 1 i.e. the dialysis in absence of magnetic field (VIAL **d**). Dialysis using slide-a-lyzer cassette with 10 kD molecular weight cut-off was performed against de-ionized water during a period of one hour. The inset of Figure 2 displays a TEM image of nanoparticle aggregates obtained in such a process. The aggregates of average diameter $D^{Sph}$ = 180 ± 10 nm were found to be dense spheres (see Supporting Information, Figure S5), indicating that they were grown by a nucleation and growth process rather than by diffusion- or reaction-limited aggregation, which should then lead to fractal clusters. Moreover, the aggregates exhibited a remarkable colloidal stability with time, since neither destabilization nor collapse of the constructs was noted over period of months. Assuming a volume fraction of 0.20 inside the large spheres [8], we have estimated that a 180 nm aggregate was built from ~ 4000 particles.

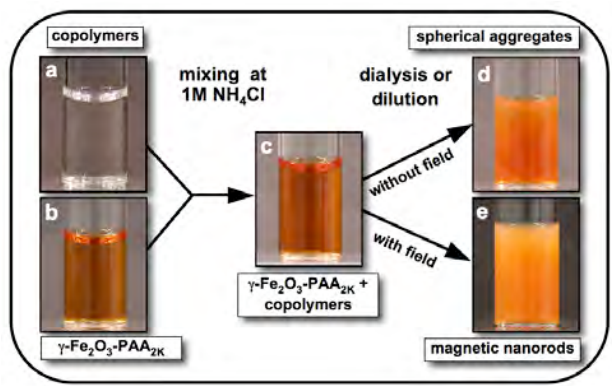

**Figure 1** : Scheme developed for the co-assembly of nanoparticle and polymer in aqueous solution. VIAL **a** : poly(trimethylammonium ethylacrylate methylsulfate-b-poly(acrylamide) in 1 M ammonium chloride ($NH_4Cl$) and weight concentration c = 0.2 wt. %. VIAL **b** : $PAA_{2K}$-coated iron oxide nanoparticles in the same ionic strength, pH and concentrations as for VIAL a. VIAL **c** : mixed solution of polymer and nanoparticles, same conditions as for VIALS **a** and **b**. The relative amount of each component was monitored by the ratio of the electrostatic charges borne by the two species. VIAL **d** : dispersion of spherical nanoparticle clusters obtained by dialysis or dilution. VIAL **e** : dispersion of superparamagnetic nanorods obtained by dialysis in presence of an external applied magnetic field.

Dilution was monitored by slow addition of de-ionized water, at a rate that was of the same order of magnitude than that of the dialysis ($dI_S/dt$ ~ $10^{-3}$ M $s^{-1}$). This second protocol was achieved in order to allow a comparison with the previous method, and to access the intermediate ionic strength values between 1 M and 10 mM. For the sake of simplicity, we focus here on the stationary states of the clustering rather than on the kinetics. This aspect will be treated separately in a forthcoming paper. As a result of slow dilution, an abrupt transition was found with decreasing ionic strength, at the critical value $I_S^0$ = 0.39 ± 0.02 M. Below $I_S^0$, large aggregates with hydrodynamic diameters $D_H^{Sph}$ = 270 ± 30 nm spontaneously formed through the association of the oppositely charged species. In terms of nanostructures, the results in Figure 2 show that dialysis and dilution are acting similarly with respect to the control of the ionic strength. It should be finally noticed that for the $PAA_{2K}$-coated nanoparticles alone (open circles), the hydrodynamic sizes remained unchanged over the whole $I_S$-range.

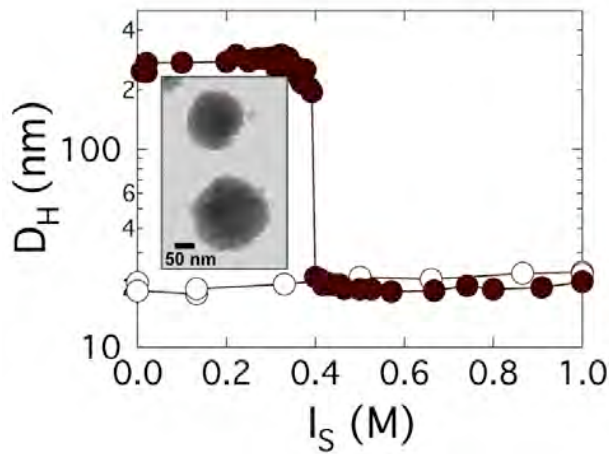

**Figure 2** : Ionic strength dependence of the hydrodynamic diameter for $\gamma$-$Fe_2O_3$-$PAA_{2K}$ nanoparticles (empty symbols) and for a dispersion containing both $\gamma$-$Fe_2O_3$-$PAA_{2K}$ particles and $PTEA_{11K}$-$b$-$PAM_{30K}$ block copolymers (closed symbols). With decreasing $I_S$, an abrupt transition was observed at the critical value $I_S^0$ = 0.39 ± 0.02 M. The transition between dispersed and clustered nanoparticles resulted from the combined processes of electrostatic complexation of the cationic blocks with the particles and of slow nucleation and growth of the aggregates. Inset : TEM image of spherical nanoparticle aggregates obtained by dialysis with average diameter $D_{Sph}$ = 180 ± 10 nm.

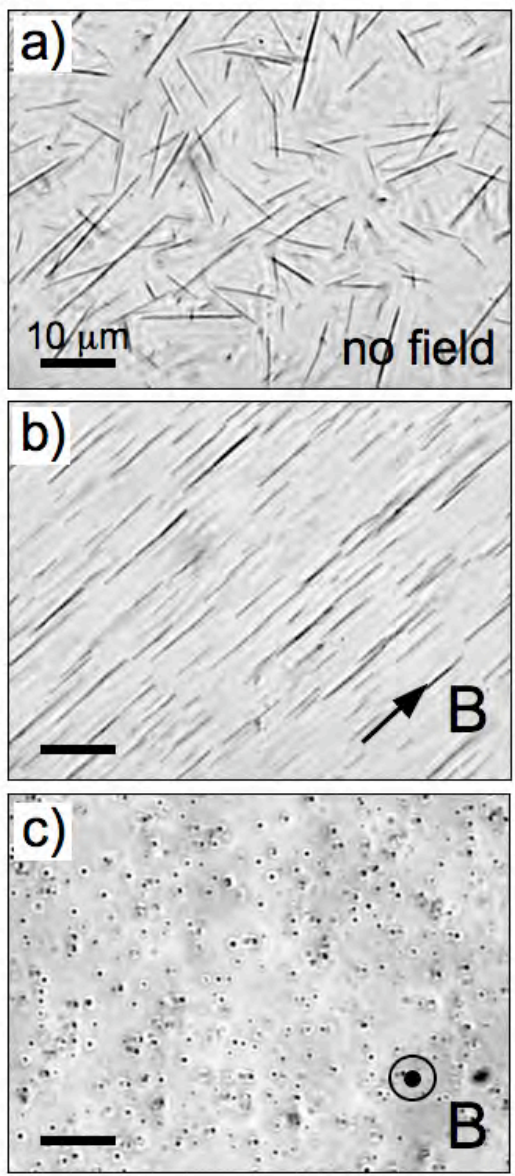

removed and the solutions were studied by optical microscopy. Figure 3a shows an image (×100) of a nanorod dispersion (VIAL **e**) sealed between glass plates. Elongated structures were clearly visible, with typical sizes in the micrometer range. For this specimen, an image analysis has allowed to derive the length distribution of the rods. The average length was estimated at 12.3 µm with a polydispersity s = 0.50 (Supporting Information, Figure S7). It is important to note that as for the spherical aggregates, the nanorods did not display signs of destabilization, even after several months. As shown in Figure 3a, the nanorods were randomly oriented in absence of magnetic field. However, when a magnet was brought near to the cell, the rods reoriented spontaneously and followed the external magnetic field lines. Figure 3b and 3c illustrate orientations in the plane of the figure and perpendicular to it, respectively. The origin of magnetic coupling is examined below. Movies of the reorienting nanorods are provided in Supporting Information.

The inset and main frame of Figure 4 exhibit TEM images of the nanostructured rods. In accordance with the optical microscopy results, linear and rigid threads were observed at the µm-scale (inset). On a shorter scale (main frame), the linear threads were found to be constituted by a multitude of 7 nm nanoparticles, presumably hold together by copolymers. In Figure 4, the diameter of the nanorods was 250 nm, a value that compared well with the diameter of the spherical aggregates obtained previously (Figure 2).

**Figure 3** : Phase contrast optical microscopy images (×100) of a nanorod dispersion (VIAL **e**) without magnetic field (a), with the magnetic field applied perpendicular (b) and parallel (c) to the optical axis of the microscope. The dots in c) display the nanorods seen from above. For this dispersion, the average length of the rods was found to be 12.3 µm, and the polydispersity s = 0.50 (Supporting Information, Figure S7). The spatial field displayed in the figures is 66×49 µm$^2$. Movies of the reorienting nanorods under changes in the orientation of the field are provided in Supporting Information.

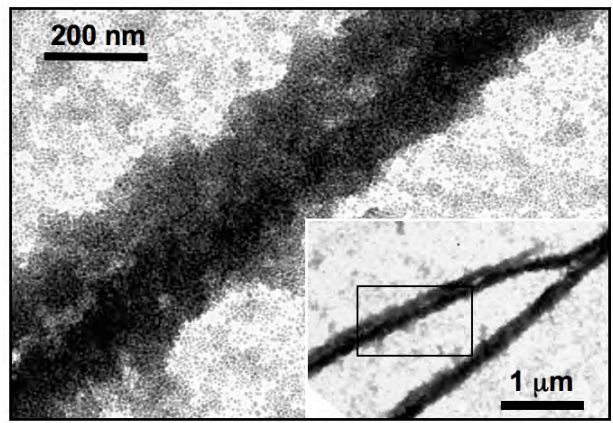

**Figure 4** : Transmission electron microscopy of a nanostructured rod constituted by a multitude of 7 nm $\gamma$-$Fe_2O_3$ nanoparticles. The diameter of the nanorods was estimated at 250 nm and the number of particles per micrometer of length at 5×10$^4$. Inset : Same nanorod at a larger length scale. The main frame covers a spatial field of 0.8×1.1 µm$_2$ and the inset of 2.6×3.7 µm$_2$.

In a third experiment, dialysis of the mixed salted solutions was operated under a constant magnetic field of 0.1 T. Once the ionic strength of the dialysis bath has reached its stationary value, the magnetic field was Concerning the mechanism of growth of the one-dimensional aggregates, we anticipate that the nanostructured rods might result from the combination of a nucleation and growth process and of the alignment of some intermediate sized-aggregates driven by the magnetic field. The data of Figure 2 also suggest that the onset of aggregation should coincide with the desorption-adsorption transition of the polyelectrolyte blocks on the charged spheres, a phenomenon largely investigated by Monte Carlo simulations [18]. Note finally that the present elongated and stiff structures bear some similarities with the filaments of magnetic microbeads recently designed for biomedical applications [19]. The main differences here are the sizes of the initial particles and the stiffness of the final constructs. Assuming a volume fraction of magnetic material inside the rods of 0.20 [8], we have estimated the linear density of particles at $5 \times 10^4$ µm$^{-1}$.

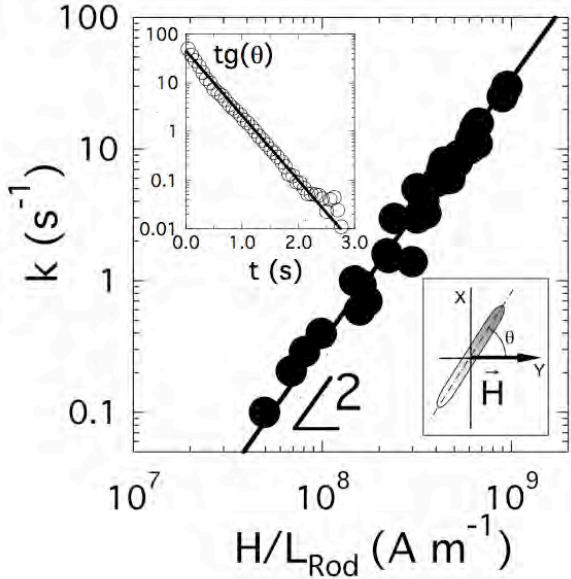

**Figure 5** : Variation of the parameter k as a function of the ratio $H/L_{Rod}$, where k is the slope of the exponential decay predicted in Eq. 2a, H the external magnetic excitation applied to the rod and $L_{Rod}$ its length. The straight line was computed according to Eq. 2b, in agreement with a quadratic dependence $k \sim (H/L_{Rod})^2$. Upper left inset : Tangent of the reorientation angle θ(t) as a function of the time. The parameter k for this set of data was k = 2.9 s$^{-1}$. Note that the exponential decay was observed over 4 decades in the ordinate units. Lower right inset : schematic representation of the reorientation of the rod initially along the X-axis and subjected to a field applied along the Y-axis.

The two prominent features displayed by the nanorods are *i*) a very large persistence length and *ii*) the ability to reorient under the external action of a magnetic field.

The persistence length is defined as the length over which the bending energy of an elongated system amounts its thermal energy. Because the threads remained stiff, even under forced rotation, we have concluded that their persistence length should be larger than their actual length (> 30 µm). More accurate estimates based on micromanipulations of the rods will be provided in a forthcoming paper. Concerning the second feature, we here demonstrate that the rods have inherited the properties of the MNPs, namely to be superparamagnetic. We have performed quantitative measurements of the kinetics associated with reorientations, and describe these results in the sequel of this communication. Let us denote by θ(t) the angle between the major axis of a rod and applied excitation **H**. The magnetic torque exerted on the rods reads $\vec{\Gamma}_{Mag} = \mu_0 V \vec{M} \wedge \vec{H}$, where $\mu_0$ is the vacuum permeability, V the volume of the rod (assimilated to an ellipsoid of major and minor axes $L_{Rod}$ and $D_{Rod}$, respectively, *i.e.* $V_{Rod} = \frac{\pi}{6} D_{Rod}^2 L_{Rod}$), $\vec{M}$ its magnetization and $\vec{H}$ the external magnetic excitation. Assuming that the rods have a magnetic susceptibility χ, the torque expresses as [20] :

$$\Gamma_{Mag} = -\frac{\mu_0 V_{Rod}}{2} \frac{\chi^2}{2+\chi} H^2 \sin(2\theta) \quad (1)$$

The quadratic dependence of the torque as a function of H is a signature of the superparamagnetic character of these aggregates. At the application of an external field at 90° with respect to the initial conditions, the nanorods rotated in the plane of observation around their center of gravity, in a propeller–like motion. The reorientational dynamics results from the balance between the magnetic torque (Eq. 1) and the hydrodynamic drag, written as $\Gamma_R = -\zeta_R d\theta/dt$ [7,20,21]. Ignoring inertial terms, the solution of the differential equation $\Gamma_R + \Gamma_{Mag} = 0$ gives :

$$\tan\theta(t) = \tan\theta_0 \exp(-kt) \quad (2a)$$

where

$$k = \frac{\chi^2}{(2+\chi)} \frac{\mu_0 g(L_{Rod}/D_{Rod})}{2\eta} D_{Rod}^2 \left(\frac{H}{L_{Rod}}\right)^2 \quad (2b)$$

In Eq. 2a, $\theta_0$ was the initial position of the rods prior to the application of the field, η is the viscosity of the suspending solvent and $g(L_{Rod}/D_{Rod})$ is a slowly varying function of the aspect ratio $L_{Rod}/D_{Rod}$ [22]. Note that in Eq. 2a, $\theta_0 \neq 90°$ since otherwise magnetic torque would be zero, and the rod would remain in an unstable equilibrium. All experiments were then performed with $\theta_0 = 80 - 87°$ using a procedure described in the Methods section. The upper right inset in Figure 5 illustrates the time dependence of tgθ(t) for a rod of length 12 µm. The semilogarithmic representation was implemented in order to emphasize the exponential character of the reorientation towards θ = 0°. The main frame in Figure 5 displays the variation of the decay rate k as a function of the ratio $H/L_{Rod}$ for 31 runs involving 12 different nanorods ($L_{Rod}$ = 2 – 30 μm) at three values of the applied field. The $(H/L_{Rod})^2$ dependence is clearly demonstrated, proving hence the validity of the assumption : the rods are superparamagnetic. From the slope of the scaling law in Figure 5, the magnetic susceptibility of this novel material was retrieved. We have found χ = 0.7. By comparison with the linear slope of the magnetization curve (Figure S3), χ = 0.7 corresponds to a local volume fraction of MNPs inside the rods of ~ 0.20 [23], a result in fair agreement with that deduced from the cryo-TEM images [8].

In conclusion, we have shown that the electrostatic interactions between oppositely charged magnetic nanoparticles and polymers can be accurately controlled by tuning the ionic strength of the dispersion. Under appropriate conditions, this control resulted in the fabrication of elongated aggregates which display remarkable stiffness and magnetic properties. These present approach opens new perspectives for the design of nanodevices such as tips, tweezers, actuators etc… applicable in biophysics and biomedecine for the stimulation and sorting of living cells or for drug delivery as novel contrast agents.

## Methods

Poly(trimethylammonium ethylacrylate)-b-poly(acrylamide) block copolymers were synthesized by MADIX® controlled radical polymerization [24,25]. The copolymer used in this work was abbreviated $PTEA_{11K}$-b-$PAM_{30K}$, where the indices indicate the molecular weights targeted by the synthesis. Light scattering performed in the dilute regime have revealed a molecular weight, $M_w^{Pol}$ = 35 000 ± 2000 g mol-1 and an hydrodynamic diameter $D_H^{Pol}$ = 11 ± 1 nm [17]. The synthesis of magnetic nanoparticles was based on the polycondensation of metallic salts in alkaline aqueous media, yielding the formation of maghemite (γ-$Fe_2O_3$) nanoparticles of sizes comprised between 4 and 15 nm [26]. The MNPs were stabilized by electrostatic interactions arising from the native cationic charges at the surface of the particles. In the conditions of synthesis (pH 1.8, weight concentration c = 6.58 wt. %), the magnetic dispersions were thermodynamically stable over a period of years. Particle sizes and size distribution were characterized by vibrating sample magnetometry, electron microscopy and magnetic sedimentation [27]. For the MNPs put under scrutiny, this distribution was described by a log-normal function with median diameter $\bar{D}_{NP}$ = 7.1 ± 0.3 nm and a polydispersity s = 0.26 ± 0.3. Electron beam microdiffraction experiments were realized in order to ascertain the crystalline structure of the maghemite (see Supporting Information, Fig. S4). In order to improve their colloidal stability, and in particular at ionic strength as large as 1 M in $NH_4Cl$, MNPs were coated with oppositely charged oligomers. We have exploited the precipitation-redispersion mechanism which was evidenced for the first time on 7 nm cerium oxide nanoparticles [28]. The precipitation of the cationic γ-$Fe_2O_3$ dispersion by oppositely charged polyelectrolytes ($PAA_{2K}$, Sigma Aldrich) was performed in acidic conditions by addition of polymers at a 1:1 weight ratio with respect to the MNPs. As the pH of the solution was increased by addition of ammonium hydroxide, the precipitate redispersed spontaneously. The hydrodynamic size of the $PAA_{2K}$-coated γ-$Fe_2O_3$ nanoparticles was found at $D_H^{NP-PAA}$ = 19 nm, i.e. 5 nm larger than the hydrodynamic diameter of the uncoated particles. This increase was interpreted as arising from the $PAA_{2K}$ brush that surrounds the nanoparticles. For the $PAA_{2K}$-coated particles, we have estimated a structural charges of -1000e per particle (where e is the elementary charge), assuming that 50 % of the 70 $PAA_{2K}$ were actually adsorbed at the iron oxide interface. For the polymers, the structural charge was assimilated to the degree of polymerization of the cationic block (+ 41e).

## Experimental

Static and dynamic light scattering have played an important role in the present work. Both techniques enabled the quantitative determination of the size, size distribution and molecular weight of the MNPs as well as to quantify the amount of $PAA_{2K}$ adsorbed on the MNPs [28]. A detailed description of this experiment can be found in the Supporting Information.

Phase-contrast images of the nanorods were acquired on an IX71 inverted microscope (Olympus, Rungis, France) equipped with a 100X objective. We used a Photometrics Cascade camera (Roper Scientific, Evry, France) and Metaview software (Universal Imaging Inc.). For the observations of the reorientations under external magnetic field, we used a Leitz (Ortholux) upward microscope with a ×20 objective. The magnetic field was applied using two pairs of coils orthogonal to each other [20]. A nanorod was selected and oriented by the first pair of coils by an AC current operating at 10 Hz (along the X-direction in lower inset of Figure 5). Once aligned, this primary field was removed and the rod was let free to move by Brownian motions for a couple of seconds. Doing so, the rods were slightly misaligned, with an angle θ0 comprised between 80° to 87°. We recall that denotes the angle between the major axis of a rod and the Y-direction. A DC current was then applied on the perpendicular coils along the Y-direction by a power-supplied generator. As a result, the rods rotated in the plane of observation from $\theta_0$ to 0°. In this configuration, the experiments were carried out at different values of the magnetic field, between $10^3$ - $10^4$ A m$^{-1}$. Video sequences were recorded by a CCD camera, digitized and treated by the ImageJ software (http://rsb.info.nih.gov/ij/). Movies of rod reorientations are shown in Supporting Information.

## Acknowledgements

We are very grateful to Ling Qi, Jean-Paul Chapel and Jean-Christophe Castaing from the Complex Fluids Laboratory (CRTB Rhodia, Bristol, Pa) for discussions and comments during the course of this study; to Christophe Lavelle and Eric Le Cam from the Institut Gustave Roussy (Université Paris-Sud) for complementary TEM experiments and helpful discussions; to Benoit Ladoux from the Laboratoire Matière et Systèmes Complexes (Université Paris-Denis Diderot) for access to microscopy and imaging facility and to Andrejs Cebers (Riga Univ., Latvia) for helpful discussions. Aude Michèle (LI2C-UPMC-Université Paris 6) is kindly acknowledged for the TEM and micro-diffraction experiments. This research is supported in part by Rhodia.